\documentclass{PoS}

\usepackage{graphicx,subfigure}
\usepackage{bm}
\usepackage{color}
\newcommand{\be}{\begin{equation}}
\newcommand{\ee}{\end{equation}}
\newcommand{\ba}{\begin{eqnarray}}
\newcommand{\ea}{\end{eqnarray}}

\newcommand{\J}{{\cal J}}

\title{Quantum field theories on the Lefschetz thimble}

\ShortTitle{Quantum field theory on a Lefschetz thimble}
\author{\speaker{M.~Cristoforetti}\\
	ECT*/FBK, strada delle tabarelle 286, 38123 Trento, Italy\\
	LISC/FBK, via Sommarive 18, 38123 Trento, Italy\\
	E-mail: \email{mcristofo@ectstar.eu}}

\author{F.~Di~Renzo\\
	Universit\`a di Parma and INFN gruppo collegato di Parma, Viale G.P. Usberti n.7/A, 43124 Parma, Italy\\
	E-mail: \email{francesco.direnzo@fis.unipr.it}}

\author{A.~Mukherjee\\
	ECT*/FBK, strada delle tabarelle 286, 38123 Trento, Italy\\
	LISC/FBK, via Sommarive 18, 38123 Trento, Italy\\
	E-mail: \email{mukherjee@ectstar.eu}}

\author{\speaker{L.~Scorzato}\\ 
       INFN-TIFPA, Trento Institute of Fundamental Physics and Applications,\\ 
       via Sommarive 24, 38123 Trento, Italy\\ 
       E-mail: \email{luigi@scorzato.it}}

\abstract{In these proceedings, we summarize the Lefschetz thimble approach to the sign problem of Quantum Field
  Theories.  In particular, we review its motivations, and we summarize the results of the application of two
  different algorithms to two test models.}

\FullConference{31st International Symposium on Lattice Field Theory - LATTICE 2013\\
		July 29 - August 3, 2013\\
		Mainz, Germany\\$\,$
		\\
		QCD-TNT-III-From quarks and gluons to hadronic matter: A bridge too far?,\\
		2-6 September, 2013\\
		European Centre for Theoretical Studies in Nuclear Physics and Related Areas (ECT*), Villazzano, Trento (Italy)}

\begin{document}

\section{Introduction}
Many important physical systems are characterized by complex actions, when formulated in terms of a path integral.
But, if the action $S$ is not real, then $e^{-S}$ is not positive semi-definite and it cannot be interpreted as a
probability distribution.  In these cases, Monte Carlo calculations are not applicable directly.  This is the so
called {\em sign problem}.  Many techniques have been proposed to overcome this problem, with important partial
successes, but the sign problem is still unsolved for a variety of important physical systems and parameter values.
In this context, any new idea that could improve our chances to simulate any of these models on larger lattices
than are feasible today would be extremely valuable.  

Recently, a new approach to control the sign problem was proposed in \cite{Cristoforetti:2012su} and further
developed in \cite{Cristoforetti:2013wha,Mukherjee:2013aga,Fujii:2013sra}.  The approach consists in reformulating
the quantum field theory (QFT) on a Lefschetz thimble \cite{Witten:2010cx,Cristoforetti:2012su}.  The Lefschetz
thimble, associated with a saddle point $\phi$, is defined as the hypersurface formed by the union of all paths of
steepest descent (SD) of the complex action ending in that saddle point $\phi$ at infinity.  Both the Lefschetz
thimble and the saddle point are constructed in an enlarged space obtained by complexifying each field component.

\section{QFT on a Letschetz  thimble}
In this section we briefly review the formulation of a quantum field on a Lefschetz thimble.  Consider a QFT on a
lattice (or any other system with a finite number of continuous degrees of freedom) defined by the action
$S(\phi)$, where $\phi$ is a vector field whose number of components, $n$, is equal to the number of degrees of
freedom in the system. Suppose that the initial field theory is defined for real fields, i.e, the expectation value
of any observable $\mathcal{O}$ is given by,
\be
\langle \mathcal{O}\rangle = \frac{\int_{\mathcal{D}} d\phi \mathcal{O} (\phi) e^{- S (\phi)}}{\int_{\mathcal{D}} d\phi e^{- S (\phi)}}
\ee
where $\mathcal{D}$ is the appropriate integration cycle for $S$ in the real domain $\mathbb{R}^n$.  Now, consider
$S$ in terms of the complexified fields, i.e, the field components $\phi_i$ are now allowed to be complex.
Suppose, $S(\phi)$ is holomorphic in this complexified space and its critical points $\phi^{\sigma}$ given by 
\be
\label{eqncr}
\frac{\partial S}{\partial \phi^{\sigma}} = 0
\ee
are non-degenerate,
\be
\det \left [ \frac{\partial^2 S}{ \partial \phi^{\sigma} \partial \phi^{\sigma}} \right ] \neq 0.
\ee 
Then, under suitable conditions on $S$ and $\mathcal{O}$ (typically fulfilled in physical systems) and for a
generic choice of parameters, we have that \cite{Witten:2010cx,Witten:2010zr,Pham:1983}
\be
\label{eqntb}
\int_{\mathcal{D}} d\phi \mathcal{O} (\phi)  e^{-S(\phi)} = \sum_{\sigma}m_\sigma\int_{\mathcal{J}_\sigma} d\phi  \mathcal{O} (\phi) e^{-S(\phi)}\;,
\ee
for some $m_\sigma\in\mathbb{Z}$.  That is, an integral over the real domain $\mathcal{D}$ is equivalent to sum of
integrals over the Lefschetz thimbles $\mathcal{J}_\sigma$.  This result can be seen as a generalization of contour
deformation in one dimension.  The Lefschetz thimbles $\mathcal{J}_{\sigma}$ associated with the critical points
are many dimensional generalizations of the paths of steepest descent. The thimble $\mathcal{J}_{\sigma}$ is
defined as the union of all paths governed by, 
\be
\label{eqnsd}
\frac{d\phi}{d\tau} = - \overline{\frac{\partial S}{\partial \phi}}
\ee
and which end at the critical point $\phi^{\sigma}$ for $\tau \to \infty$. They are hypersurfaces of \emph{real}
dimension $n$ embedded in the complex manifold $\mathbb{C}^n$. Here, and later, the overhead bar represents complex
conjugation In this paper we will assume that $S$ is a Morse function, i.e., it has only non degenerate critical
points\footnote{Degenerate minima, as they typically occur in the presence of symmetries, can be either lifted or
  treated as discussed in \cite{Cristoforetti:2012su}.}.

On a Lefschetz thimble $\mathcal{J}_{\sigma}$, the imaginary part of the action $\Im{S(\phi)}$ remains constant.
The measure term does introduce a new {\em residual phase}, due to the curvature of the thimble, but it is expected
to be a smooth function of the fields (see later).

Eq.~(\ref{eqntb}) represents an exact reformulation of the original path integral, but, for a QFT, reproducing the
original integral exactly as in Eq.~(\ref{eqntb}) is both impractical and unnecessary.  In fact, for a large class
of QFTs one can show that the regularization on a (suitably chosen) single thimble defines a QFT with the same
degrees of freedom, the same symmetries (and symmetry-representations), the same perturbative expansion and the
same naive continuum limit as the traditional formulation\footnote{In \cite{Cristoforetti:2012su} this was shown
  explicitely only for QCD and for a self-interacting complex scalar field, but the arguments are much more
  general.}.  On the basis of {\em universality}, these are the properties that justify a legitimate {\em
  regularization} of that QFT, and the formulation on a single thimble is expected to coincide with the traditional
formulation, after suitable renormalization and in the continuum limit\footnote{It would be also desirable to
  understand the role of reflection positivity in this setup.  This is not easy, because even in the ordinary
  formulation naive reflection positivity does not hold, and it is not clear how it should be reformulated.
  Hopefully the analysis started in \cite{Jaffe:2012nc,Jaffe:2013yia} will shed some light on this issue.}. (The
same argument can be applied also to statistical models, at least in the vicinity of critical points.)

This leads to the formulation on a single thimble $\mathcal{J}_0$ as: 
\be
\label{eq:obs_thi}
\langle \mathcal{O} (\phi) \rangle =\frac{ \int_{\mathcal{J}_0} d\phi \mathcal{O} (\phi) e^{- S (\phi)}}{\int_{\mathcal{J}_0}  d\phi (\phi) e^{- S (\phi)}} \;,
\ee

Note that universality is not a theorem (beyond perturbation theory), but it is nevertheless a property that one
needs to assume anyway when studying any QFT on the lattice.  So, this approach does not require {\em new}
assumptions.  It might certainly be that our present understanding of universality is incomplete, and the
properties listed above do not define a QFT, not even in the continuum limit.  But the thimble approach offers a
well defined setup to investigate also this very fundamental question, by comparing the results obtained on
different thimbles (although reproducing precisely that combination of thimbles that corresponds exactly to the
traditional formulation seems not feasible).

Ref.~\cite{Cristoforetti:2012su} put forward another argument to justify the formulation on a single thimble (see
also \cite{Witten:2010cx}).  In fact, it is often possible to identify a single stationary point such that the
contribution of the thimble attached to it is dominant, in the sense that the contribution of the other thimbles is
either zero or exponentially suppressed, in presence of a large number of degrees of freedom.  This may happen
either because the system has a single global minimum, which is also a stationary point of the complexified action,
or because there are degenerate global minima, that are however connected by symmetries, or because there are
degenerate global minima with vanishing probability of tunneling.  (When the global minimum is not a stationary
point of the complexified action, it is always easy to locate the stationary point whose thimble contains the
global minimum.  This is typically the dominant one.)

Note that this latter argument is not conclusive: it is possible, e.g., that a large number of stationary points
may still accumulate near the global minimum giving a finite contribution.  But, if universality is correct, these
other contributions should either be {\em equivalent}, or they should represent {\em lattice artifacts}, that are
effectively removed by the renormalization procedure.

There is certainly a deep connection between the Lefschetz thimble and the complex Langevin
\cite{Bongiovanni:2013nxa} approaches.  At the very least, because both share the same stationary points, although
the relevant phase spaces are very different.  The advantage of our approach is that a Monte Carlo on a thimble
provably simulates a well defined functional integral, which, in turns, is related to the usual formulation on the
basis of standard assumptions.  We refer to
\cite{Ferrante:2013hg,Pehlevan:2007eq,Guralnik:2007rx,Basar:2013eka,Aarts:2013fpa,Eruzzi2013} for further insight
in this topic.

\vskip 0.5 cm

Different numerical approach can be envisaged to perform the integration on the Lefschetz thimble. In the following
we present the results obtained testing two of them: for the $U(1)$ one-link model we apply a Metropolis algorithm
while for the relativistic Bose gas we use the Langevin dynamics. Details on the implementation of the two methods
are presented in \cite{Cristoforetti:2013wha,Mukherjee:2013aga}. (See \cite{Fujii:2013sra} for another possible
algorithm.)

\section{One-plaquette model with $U(1)$ symmetry}
As a first simple application of the method we now discuss a system 
with one degree of freedom, viz. the one-plaquette model with $U(1)$
symmetry.
The action is given in terms of the gauge link $U=e^{i\phi}$ as
\be
S=-i\frac{\beta}{2}\left(U+U^{-1}\right)=-i\beta\cos\phi .
\ee
where $\phi$ in this case is a one component field.
For real $\beta$ the action is complex, similar to real time gauge theories.

For this simple model, all the integrals can be evaluated analytically, which offers the chance to compare every detail of 
our numerical results to exact results. In particular the plaquette average of the phase $e^{i\phi}$ is given by,
\be
\langle e^{i\phi} \rangle = i\frac{J_1(\beta)}{J_0(\beta)} 
\ee
with $J_n(\beta)$ being Bessel functions of the first kind.
This analytic result offers the chance of a clear test of our algorithm.

Obtaining this result using stochastic methods is quite non-trivial.  
For example the complex Langevin method without \emph{ad-hoc} optimizations gives the wrong result for this model \cite{Berges:2007nr}.

In order to apply our method, we treat the field $\phi$ as complex. The action $S$ has two critical points at $\phi=0$ and $\pi$.
By explicitly constructing the Hessian, it is easy to show that both the critical points are non-degenerate.
In this simple model we can also compute the intersection numbers ($m_{\sigma}$), which turn out to be equal to 1 for both thimbles.
The field configurations on the two thimbles are related by the discrete symmetry transformation $\phi \to \pi - \overline{\phi}$,
and expectation values of observables can be written in terms of integrals over one thimble only. 

\begin{figure}
\subfigure{\includegraphics[width=0.5\columnwidth]{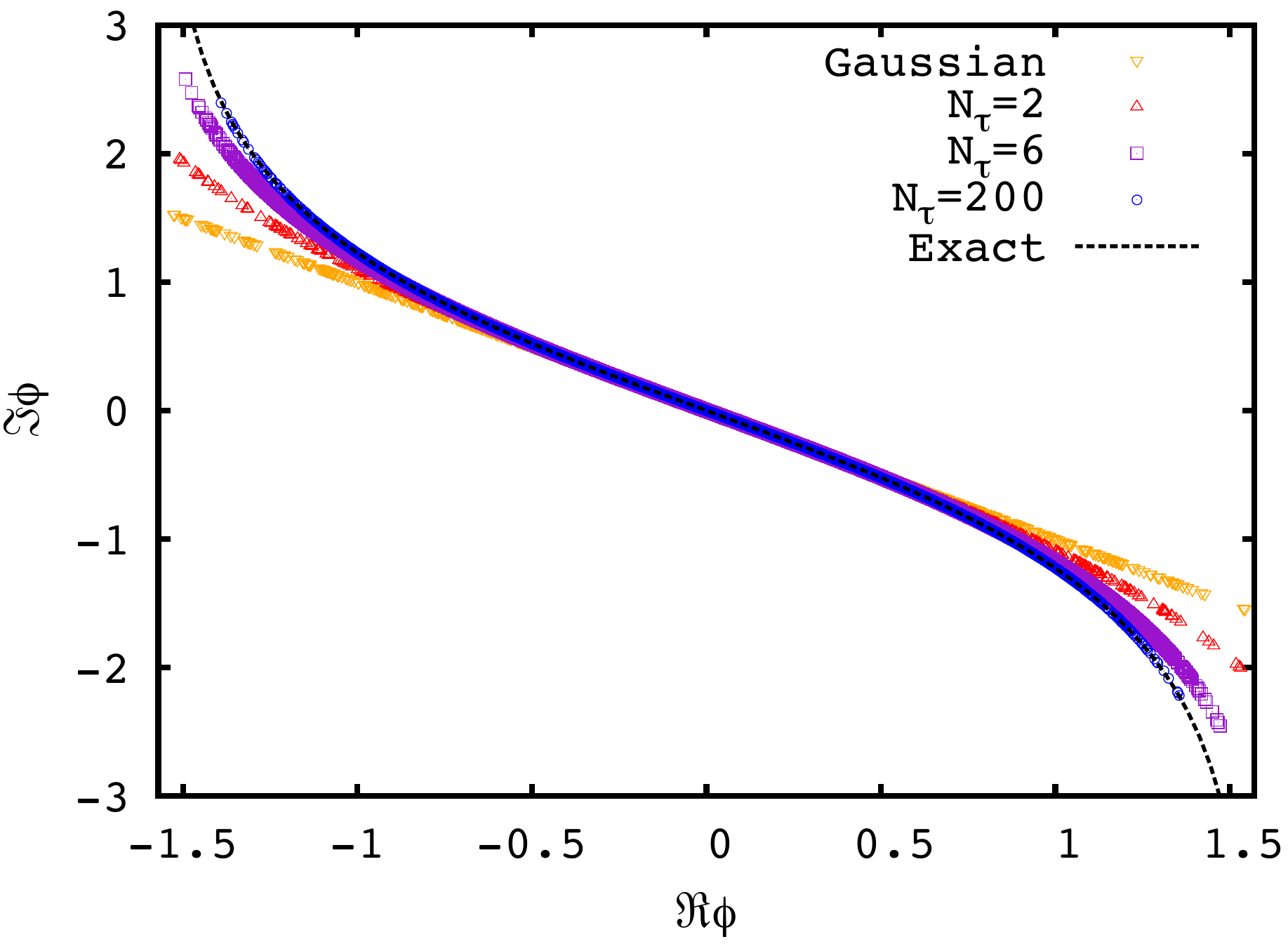}}
\subfigure{\includegraphics[width=0.5\columnwidth]{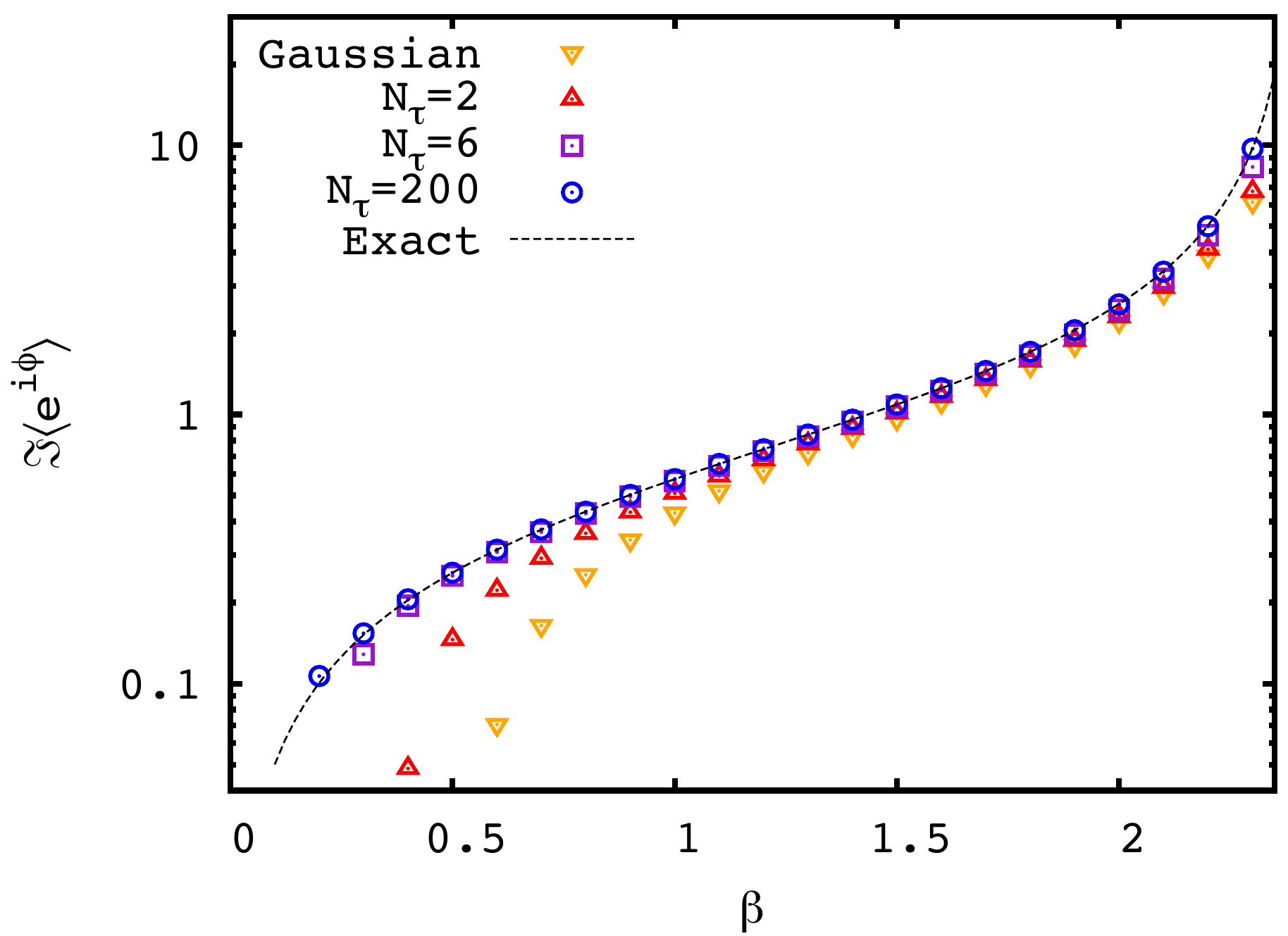}}
\caption{Left: sampled field configurations at $\beta=1$ for the thimble attached to $\phi=0$. Right: Expectation value of $e^{i\phi}$ as a function of $\beta$.}
\label{fig:thimble_conv}
\end{figure}

For this model, one can explicitly derive the expression for the thimbles attached to the two saddle points.
In the left panel of Fig.~\ref{fig:thimble_conv}  we show that the  fields obtained using the numerical method
 described above reproduce well the exact thimble.

We see systematic improvement in our results on increasing $N_{\tau} = \epsilon^{-1}$, which represent the discretization parameter in our Metropolis algorithm (the continuum limit is given by $\epsilon\rightarrow 0$. With increasing $N_{\tau}$ the sampled field configurations uniformly converge on the true thimble. 
In contrast, the flat Gaussian thimble ($N_{\tau} =1$) approximates the thimble quite well near the saddle 
point, but it noticibly different further away from the saddle point.

In the right panel of Fig.~\ref{fig:thimble_conv}  we show the results for the expectation value of the observable $e^{i\phi}$ for different $\beta$. 
Again, the results from our method systematically approach the exact analytical result with increasing $N_{\tau}$.
For $N_{\tau}=200$, the results from our method are identical 
(within statistical errors) to the analytical results for the range of $\beta$ considered. In contrast, we notice that 
there is a large difference between the analytical result and those from Monte Carlo if the field configurations are
sampled from the flat Gaussian thimble.

Finally, we discuss the residual phase in the context of the $U(1)$ one-plaquette model.
The question of the residual phase is an important one. 
We expect it to produce a milder sign problem (if at all),
than the original sign problem. Nevertheless, it should be included in any quantitative estimate. 
\begin{figure}
\centering
\includegraphics[width=.6\columnwidth]{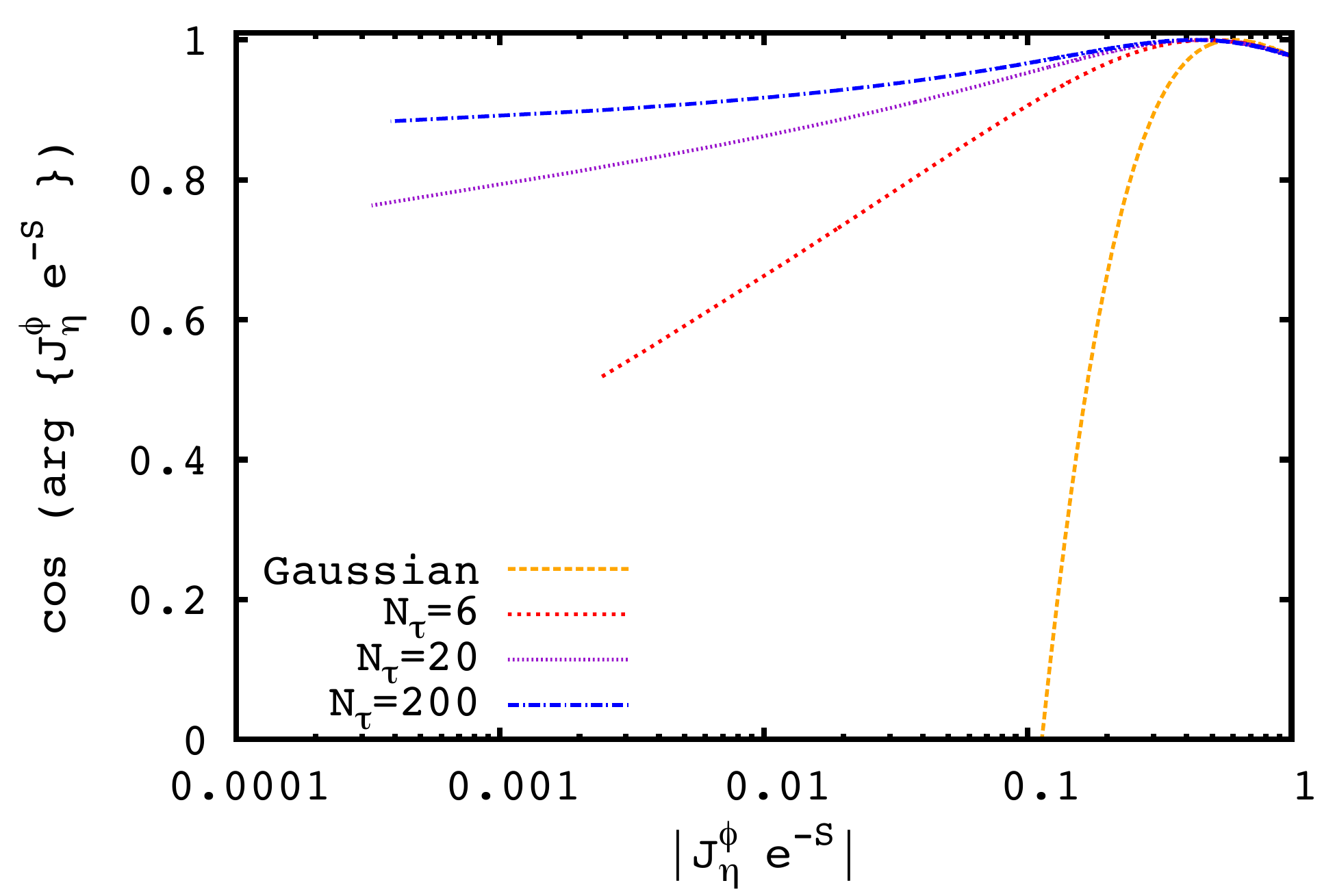}
\caption{The residual phase as a function of the probability measure at $\beta=1$.}
\label{fig:resphase}
\end{figure}
In our formulation the full (complex) measure of integration is given by $\det \left [ \textrm{J}^{\phi}_{\eta}\right ] e^{-S} $. 
The full integrals on the Lefschetz thimble are always real. This means that
$\sin \left (\arg \left \{\det \left [ \textrm{J}^{\phi}_{\eta} \right ]  e^{-S} \right \} \right ) $ does not contribute to the integral.
The statement that the sign problem in our method is mild (or absent) means that $\cos \left (\arg \left \{ \det \left [ \textrm{J}^{\phi}_{\eta} \right ] e^{-S} \right \}\right ) $ (residual phase)
will vary very little (or not at all), in the region where $\left | \det \left [\textrm{J}^{\phi}_{\eta} \right ] e^{-S} \right | $ (probability measure) is significant.

For the $U(1)$ one-plaquette model, the Jacobian of the transformation on each thimble is a single number and is simply given by,
\be
J^{\phi}_{\eta} =  \frac{-i\beta \, \overline{\sin{\phi}}}{\eta}.
\ee
In Fig.~\ref{fig:resphase} we show the residual phase vs the positive probability measure for this model.
 We see that the residual phase changes by very little 
 for variations of the probability measure spanning many orders of magnitude. Moreover, the fluctuations
of the residual phase grow milder as the true thimble is approached starting from the Gaussian thimble.
Most importantly, the residual phase keeps the same sign throughout the full domain of integration, i.e., 
there is \emph{no sign problem} for our method for this particular model.

\section{Relativistic Bose gas at finite chemical potential} 
The model is defined by the following continuum action:
\begin{equation}
\label{eq:Scont}
S= \int d^4 x[|\partial \phi|^2+(m^2-\mu^2) |\phi|^2+\mu j_0+\lambda |\phi|^4],
\end{equation}
where $\phi(x)$ is a complex scalar field, $j_{\nu} := \phi^*\partial_{\nu} \phi - \phi\partial_{\nu} \phi^*$ and
$\mu$ is the chemical potential.  In this model (as in QCD) the density $\langle n\rangle=\frac{1}{V}\partial \ln Z
/ \partial\mu$ is expected to be zero up to a critical point.  But, this phase transition is hidden in the standard
Monte Carlo method because of the strong sign problem which appears as soon as $\mu \neq 0$.

To formulate and simulate the relativistic Bose gas on a Lefschetz thimble \cite{Cristoforetti:2012su}, we need to
discretize the system defined by Eq.~(\ref{eq:Scont}) and extend the action $S$ holomorphically. This is done by
complexifying both the real and imaginary part of the original complex fields $\phi_x =
\frac{1}{\sqrt{2}}(\phi_{1,x} + i \phi_{2,x})$, as $\phi_{a,x} = \phi^{(R)}_{a,x} + i \phi^{(I)}_{a,x}$, $a=1,2$,
which leads to the action in $d$ dimensions \cite{Aarts:2009hn}:
\begin{eqnarray}
S[\{\phi_{a,x}\}]=
&&
\sum_x
  \left[
    \left(d+\frac{m^2}{2}\right)\sum_a \phi_{a,x}^2+\frac{\lambda}{4}(\sum_a \phi_{a,x}^2)^2\right.\left.
    -\sum_{a}\sum_{\nu=1}^{d-1}\phi_{a,x}\phi_{a,x+\hat{\nu}}+\sum_{a,b} i\sinh\mu\, \varepsilon_{ab}\phi_{a,x}\phi_{b,x+\hat{0}}
\right.\nonumber\\
&&\left. - \cosh\mu\, \delta_{a,b} \phi_{a,x}\phi_{b,x+\hat{0}} \right],
\label{eq:S-hol}
\end{eqnarray}
($\varepsilon$ is the 2 dimensional anti-symmetric Levi-Civita symbol).  The observables are defined as:
\begin{eqnarray}
\label{eq:Z}
\langle {\cal O} \rangle_0 = \frac{1}{Z_0} \int_{\J_0} \; \prod_{a,x} d\phi_{a,x} \; e^{-S[\phi]} {\cal
  O}[\phi], \nonumber &\qquad&Z_0 = \int_{\J_0} \; \prod_{a,x} d\phi_{a,x} \; e^{-S[\phi]},
\end{eqnarray}
where the integration domain $\J_0$ is the Lefschetz thimble \cite{Pham:1983,Witten:2010cx} attached to $\phi_{\rm
  glob}$. The configuration $\phi_{\rm glob}$ is the global minimum of the real part of the action $S_R=\Re \{S\}$,
when restricted to the original domain $\mathbb{R}^{2V}$.  More precisely, $\J_0$ is the manifold of real dimension
$N=2V$, defined as union of all the curves of SD for $S_R$, i.e., the curves that are solutions of Eq.~\ref{eqnsd}
and that end in $\phi_{\rm glob}$ for $\tau \to \infty$.

In presence of spontaneous symmetry breaking (SSB), the global minimum $\phi_{\rm glob}$ is degenerate.  But, the
whole procedure can be defined by introducing an explicit term of symmetry breaking: $h \sum_{x,a} \phi_{x,a}$,
where $h$ is a real constant, that selects a specific minimum \cite{Cristoforetti:2012uv} (that can be computed
also analytically).  Since $h$ is real, the global minimum $\phi_{\rm glob}$ of $S_R$ is also a stationary point of
the imaginary part of the action $S_I$, and hence the thimble is well defined.  Physical results are obtained by
extrapolating to $h \to 0$.

In this case we use the Langevin dynamics algorithm in order to move on the thimble, and in particular we restrict the simulation to the case of the flat Gaussian manifold ${\cal G}_0$ which in this case is a good approximation of the exact thimble (see \cite{Cristoforetti:2012su, Cristoforetti:2012uv} for details)
 The results obtained in this way agree perfectly (within the rather small errors) with the results obtained with the algorithm of \cite{Gattringer:2012df} and \cite{Aarts:2008wh}\footnote{We thank Gert Aarts, Christof Gattringer and Thomas Kloiber for sharing their  partially unpublished results with us.}.  In particular, they show the correct scaling with the volume.  

\begin{figure}
\subfigure{\includegraphics[width=.5\columnwidth]{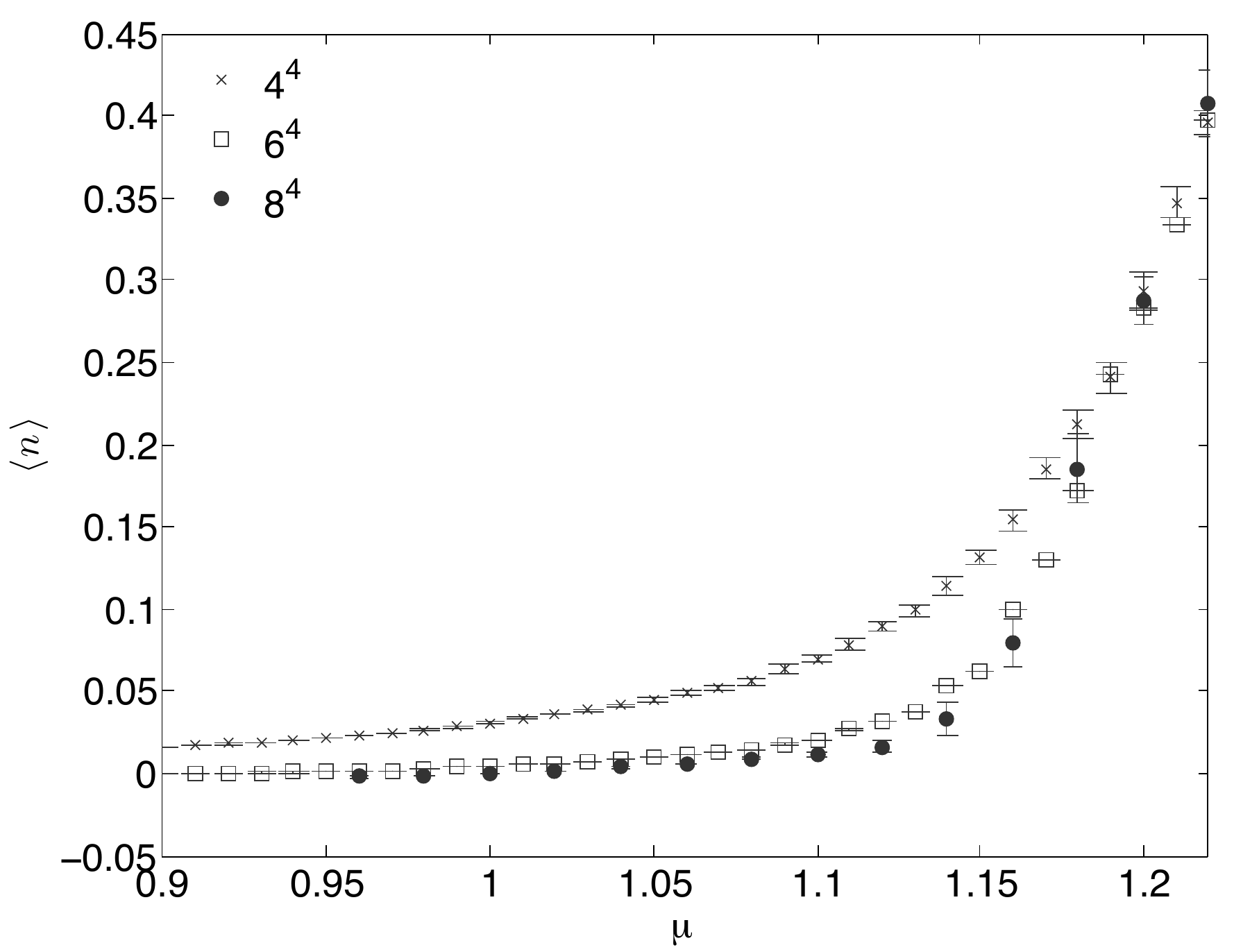}}
\subfigure{\includegraphics[width=.5\columnwidth]{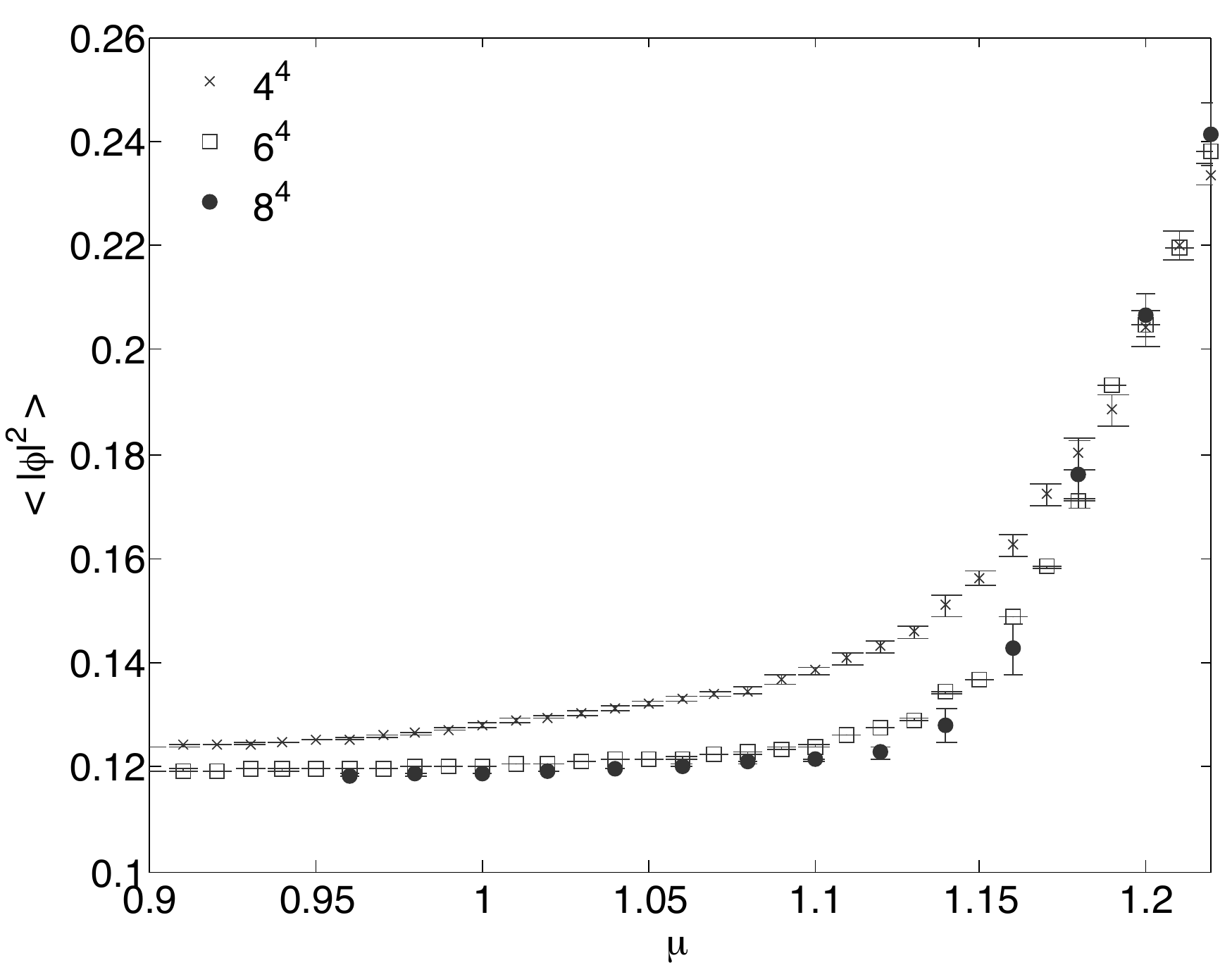}}
\label{fig:phi4_1}
\caption{Average density $\langle n\rangle$ (left) and $\langle |\phi|^2 \rangle$ (right) in the critical region for the lattices $V=4^4, 6^4,
  8^4$. Right}
\end{figure}

We report the results of simulations for the relativistic Bose gas in $3+1$ dimensions ($d = 4$). The mass and
coupling were fixed at $m = 1 = \lambda$, and $\mu$ was varied from $0$ to $1.3$.  In Fig.~\ref{fig:phi4_1} we plot our results for the density $\langle n\rangle$ and $\langle |\phi|^2 \rangle$ in the most
interesting range between $\mu=0.9$ and $\mu=1.22$.  In these figures, we see a clear signal of the Silver Blaze
phenomenon around $\mu \sim 1.1$.  In all the simulations shown here we used $\Delta t=10^{-4}$, but we performed
also some tests with $\Delta t=10^{-3}$ and $\Delta t=10^{-5}$ and we found no significant difference.  The
errorbars on each point are computed from the standard deviation of 10 to 20 independent histories, in order to
take the autocorrelation effects into account.  We used the sources $h=5\times 10^{-3}$ and $h=10^{-3}$ to extract
the limit $h\to 0$.

\begin{figure}
\centering
\includegraphics[width=.6\columnwidth]{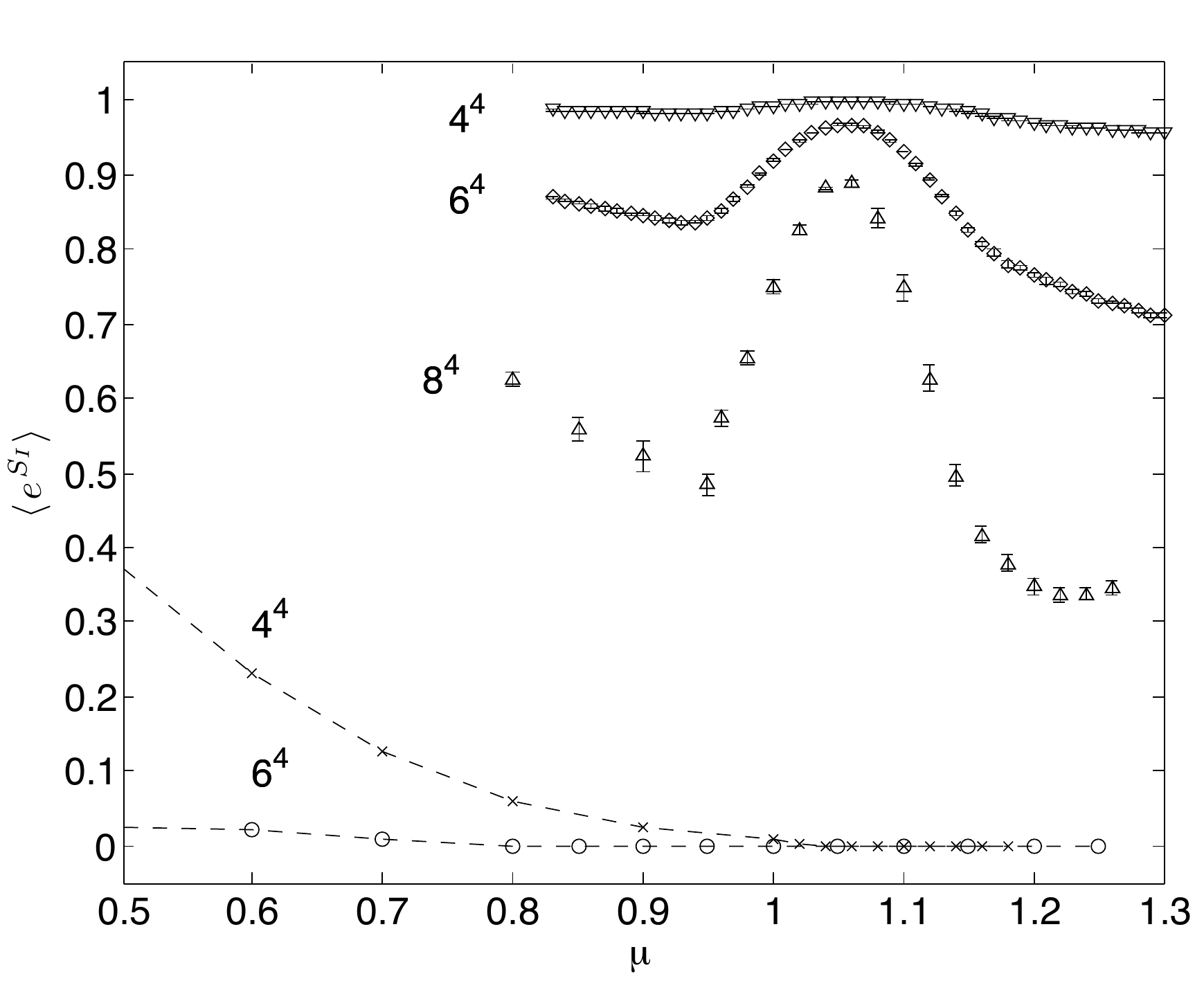}
\caption{\label{fig:fase} The data on the top-right show the average phase obtained with the Aurora algorithm on
  lattices $4^4$, $6^4$ and $8^4$.  It is interesting that the average phase is large precisely in the most
  interesting region just above $\mu=1$.  The dashed lines on the bottom-left display, for comparison, the average
  phase obtained with a naive phase-quenched Monte Carlo algorithm on lattices $4^4$ and $6^4$.  Even on a $4^4$
  lattice, the sign problem in the phase-quenched algorithm, completely hides the interesting region.}
\end{figure}

In Fig.~\ref{fig:fase} we plot the average phase for the same simulations reported above.  The phase is used to
reweight the observables.  However, such reweighting brings corrections to the observables that are unnoticeable,
within the statistical errors.  As expected, the sign problem in ${\cal G}_0$ gradually increases on larger volumes
and moving closer to the thimble will be eventually necessary.
 
Recently, a simulation on the exact thimble was carried out which verified our results and also showed that there
is no sign problem coming from the residual phase in this case \cite{Fujii:2013sra}.

\section{Conclusions}

We have reported on the status of the Lefschetz thimble approach to the sign problem of QFTs.  In particular, we
have reviewed the justification of the approach, and we have summarized its application to two different simple
models, in which we have employed two different algorithms.  The results are very encouraging and much work is
currently ongoing to design better algorithms and apply these ideas to more QFTs.

\acknowledgments 
We would like to thank Giovanni Eruzzi, Christian Torrero and Christian Schmidt for useful discussions.  LS and FDR
acknowledge support from the INFN - SUMA, the EU STRONGnet and the INFN QCDLAT projects.


\begin{thebibliography}{99}
\bibitem{Cristoforetti:2012su} 
M.~Cristoforetti, F.~Di~Renzo and L.~Scorzato, \emph{New approach to the sign problem in quantum field theories: High density QCD on a Lefschetz thimble}, Phys.~Rev.~D {\bf 86}  (2012) 074506, [{\tt arXiv:1205.3996}].

\bibitem{Cristoforetti:2013wha}
M.~Cristoforetti, F.~Di~Renzo, A.~Mukherjee and L.~Scorzato, \emph{Monte Carlo simulations on the Lefschetz thimble: taming the sign problem}, Phys.~Rev.~D {\bf 88}  (2013) 051501, [{\tt arXiv:1303.7204}].

\bibitem{Mukherjee:2013aga}
A.~Mukherjee, M.~Cristoforetti and L.~Scorzato, \emph{Metropolis Monte Carlo on the Lefschetz thimble: application to a one-plaquette model}, Phys.~Rev.~D {\bf 88}  (2013) 051502, [{\tt arXiv:1308.0233}].

\bibitem{Fujii:2013sra}
H.~Fujii, D.~Honda, M.~Kato, Y.~Kikukawa, S.~Komatsu and others, \emph{Hybrid Monte Carlo on Lefschetz thimbles - A study of the residual sign problem}, \emph{JHEP}  {\bf 1310} (2013) 147,  [{\tt arXiv:1309.4371}].

\bibitem{Witten:2010cx}
E.~Witten, \emph{Analytic Continuation Of Chern-Simons Theory},  (2010) {\tt arXiv:1001.2933}.

\bibitem{Jaffe:2012nc}
A.~Jaffe, C.~D.~Jakel and R.~E.~Marinez II, \emph{Complex Classical Fields: A Framework for Reflection Positivity}, (2012) {\tt arXiv:1201.6003}.

\bibitem{Jaffe:2013yia}
A.~Jaffe, C.~D.~Jakel and R.~E.~Marinez II, \emph{Complex Classical Fields and Partial Wick Rotations}, (2013) {\tt arXiv:1302.5935}.

\bibitem{Bongiovanni:2013nxa}
L.~Bongiovanni, G.~Aarts, E.~Seiler,D.~Sexty and I-O.~Stamatescu, \emph{Adaptive gauge cooling for complex Langevin dynamics}, (2013) {\tt arXiv:1311.1056}.

\bibitem{Ferrante:2013hg}
D.~D.~Ferrante,G.~S.~Guralnik, Z.~Guralnik and C.~Pehlevan, \emph{Complex Path Integrals and the Space of Theories}, (2013)  {\tt arXiv:1301.4233}.

\bibitem{Pehlevan:2007eq}
C.~Pehlevan and G.~Guralnik, \emph{Complex Langevin Equations and Schwinger-Dyson Equations}, Nucl.~Phys.~B {\bf 811} (2009) 519,  [{\tt arXiv:0710.3756}].

\bibitem{Guralnik:2007rx}
G.~Guralnik and Z.~Guralnik, \emph{Complexified path integrals and the phases of quantum field theory}, Annals~Phys.~325 (2010) 2486, [{\tt arXiv:0710.1256}].

\bibitem{Basar:2013eka}
G.~Basar, G.~V.~Dunne and M.~Unsal, \emph{Resurgence theory, ghost-instantons, and analytic continuation of path integrals}, \emph{JHEP} {\bf 1310} (2013) 041, [{\tt arXiv:1308.1108}].

\bibitem{Aarts:2013fpa}
G.~Aarts, \emph{Lefschetz thimbles and stochastic quantisation: Complex actions in the complex plane}, (2013)  {\tt arXiv:1308.4811}.

\bibitem{Eruzzi2013} 
G.~Eruzzi et. al. in preparation

\bibitem{Witten:2010zr}
E.~Witten, \emph{A New Look At The Path Integral Of Quantum Mechanics}, (2010) {\tt arXiv:1009.6032}.

\bibitem{Pham:1983}
F.~Pham, \emph{Vanishing homologies and the n variable saddlepoint method}, Proc.~Symp.~Pure~Math. {\bf 40} (1983) 319.

\bibitem{Berges:2007nr}
J.~Berges and D.~Sexty, \emph{Real-time gauge theory simulations from stochastic quantization with optimized updating}, Nucl.~Phys.~B {\bf 799} (2008) 306, [{\tt arXiv:0708.0779}].

\bibitem{Aarts:2009hn}
G.~Aarts, \emph{Complex Langevin dynamics at finite chemical potential: Mean field analysis in the relativistic Bose gas}, \emph{JHEP} {\bf 0905} (2009) 052, [{\tt arXiv:0902.4686}].

\bibitem{Cristoforetti:2012uv}
M.~Cristoforetti, F.~Di~Renzo and L.~Scorzato, \emph{The sign problem and the Lefschetz thimble}, J.~Phys.~Conf.~Ser. {\bf 432} (2013) 012025, [{\tt arXiv:1210.8026}].

\bibitem{Gattringer:2012df} 
C.~Gattringer and T.~Kloiber, \emph{Lattice study of the Silver Blaze phenomenon for a charged scalar $\phi^4$ field}, Nucl.~Phys.~B {\bf 869} (2013) 56, [{\tt arXiv:1206.2954}].

\bibitem{Aarts:2008wh}
G.~Aarts, \emph{Can stochastic quantization evade the sign problem? The relativistic Bose gas at finite chemical potential}, Phys.~Rev.~Lett. {\bf 102} (2009) 131601, [{\tt arXiv:0810.2089}].

\end{thebibliography}
\end{document}